\documentclass[sigconf]{acmart}

\usepackage[show]{chato-notes}
\usepackage{float} 

\usepackage{amssymb}
\usepackage{pifont}%
\newcommand{\cmark}{\ding{51}}%
\newcommand{\xmark}{\ding{55}}%

\newcommand{\sm}[1]{#1}

\acmSubmissionID{504}

\AtBeginDocument{%
  }

\copyrightyear{2026}
\acmYear{2026}
\setcopyright{cc}
\setcctype{by}
\acmConference[SIGIR '26]{Proceedings of the 49th International ACM SIGIR Conference on Research and Development in Information Retrieval}{July 20--24, 2026}{Melbourne, VIC, Australia}
\acmBooktitle{Proceedings of the 49th International ACM SIGIR Conference on Research and Development in Information Retrieval (SIGIR '26), July 20--24, 2026, Melbourne, VIC, Australia}
\acmDOI{10.1145/3805712.3809980}
\acmISBN{979-8-4007-2599-9/2026/07}

\begin{document}

\title{The Matryoshka Hypencoder}

\author{Majd Alkawaas}
\affiliation{%
  \institution{University of Glasgow}
  \city{Glasgow}
  \country{United Kingdom}
}
\email{Majd.Alkawaas@glasgow.ac.uk}
\orcid{0009-0005-8295-4354}

\author{Sean MacAvaney}
\affiliation{%
  \institution{University of Glasgow}
  \city{Glasgow}
  \country{United Kingdom}}
\orcid{0000-0002-8914-2659}
\email{Sean.MacAvaney@glasgow.ac.uk}

\begin{abstract}
The Hypencoder is a recently-proposed retrieval approach that encodes queries as shallow neural networks (``Q-Nets'') that estimate relevance over pre-computed document embeddings. Inspired by Matryoshka Representation Learning, we show that the Hypencoder can be extended to support multiple sizes of Q-Nets, allowing trade-offs between effectiveness and efficiency when deployed. We find that this ``Matryoshka Hypencoder'' achieves comparable in-domain effectiveness with approximately $7\times$ fewer active parameters in-domain and half as many active parameters out-of-domain\sm{, which corresponds to a 1.6--3.4$\times$ increase in scoring throughput. This work paves the way for practical deployment of Hypencoders.}

\vspace{0.4em}
\hspace{0.8em}\includegraphics[width=1.25em,height=1.25em]{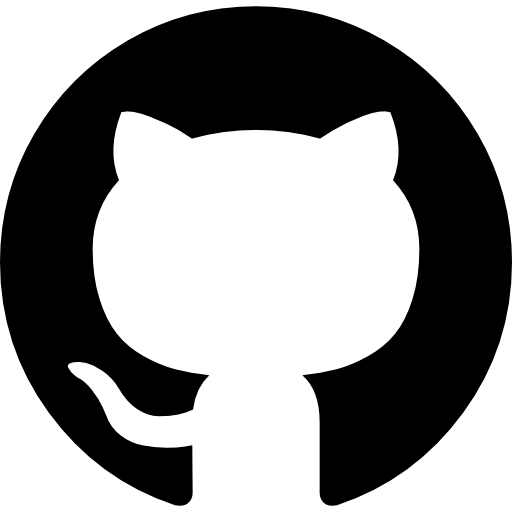}\hspace{.3em}
\parbox[c]{\columnwidth}
{
    \vspace{-.55em}
    \href{https://github.com/MajdAlkawaas/hypencoder-paper}{\nolinkurl{https://github.com/MajdAlkawaas/hypencoder-paper}}
}
\vspace{-1.2em}

\end{abstract}



\maketitle
\section{Introduction}

Neural ranking architectures provide inherent trade-offs in effectiveness, inference cost, and efficiency. For instance, single-vector dense architectures, such as DPR~\cite{DBLP:conf/emnlp/KarpukhinOMLWEC20}, have low storage and inference costs, at the \sm{expense} of inherent limitations in effectiveness~\cite{DBLP:journals/corr/abs-2508-21038}. Other architectures, such as those used by late interaction models~\cite{DBLP:conf/sigir/KhattabZ20} and cross-encoders~\cite{DBLP:journals/corr/abs-1901-04085}, sacrifice inference \sm{speed for an increased capability} to estimate relevance. \citet{hypencoder} recently proposed the \textit{Hypencoder}, which aims to give single-vector dense models a higher capability to estimate relevance by scoring document vectors using a lightweight query-specific neural network (a \textit{Q-Net}), rather than typical measures of vector similarity (e.g., cosine, inner product). The Q-Net (generated by a separate neural model, i.e., a \textit{HyperNetwork}~\cite{hypernetworks}) has greater expressive capacity than traditional similarity measures, and \sm{consequently} provides higher \sm{retrieval} effectiveness. \sm{The Hypencoder manages} storage costs by leveraging single-vector document encodings, and inference efficiency is managed by using a shallow multi-layer perceptron as the Q-Net \sm{architecture}.

Meanwhile, several works have explored alternative ways to balance trade-offs in single-vector dense models. \sm{These efforts usually assume a fixed model, instead focusing on how to efficiently index and search for a given model~\cite{DBLP:journals/pami/MalkovY20,DBLP:conf/sigir/KulkarniMGF23,DBLP:conf/icml/GuoSLGSCK20,DBLP:conf/sigir/Macdonald26jpqrepro}.} A notable exception is \citet{mrl}, which proposed learning a single model capable of representing query and document vectors at multiple sizes. These \textit{Matryoshka} models allow a single model to be deployed at different operational points---a greater number of dimensions generally improves effectiveness at the expense of greater storage and inference costs. We refer to this design as the \textit{Matryoshka Principle}: representations are constructed as nested prefixes, such that any prefix constitutes a valid, lower-cost representation, while longer prefixes strictly increase capacity and effectiveness at the expense of additional computation or storage. The Matryoshka Principle has also been applied \textit{inside} encoder transformer networks~\cite{DBLP:conf/iclr/LiLL0L25,DBLP:conf/sigir/0032ZKZ25}, allowing efficiency-effectiveness trade-offs to be made during the process of encoding queries and documents. In this case, a greater number of activated parameters (here, transformer layers) results in higher effectiveness at the cost of additional inference costs.

\begin{figure}
\centering
\includegraphics[width=\linewidth]{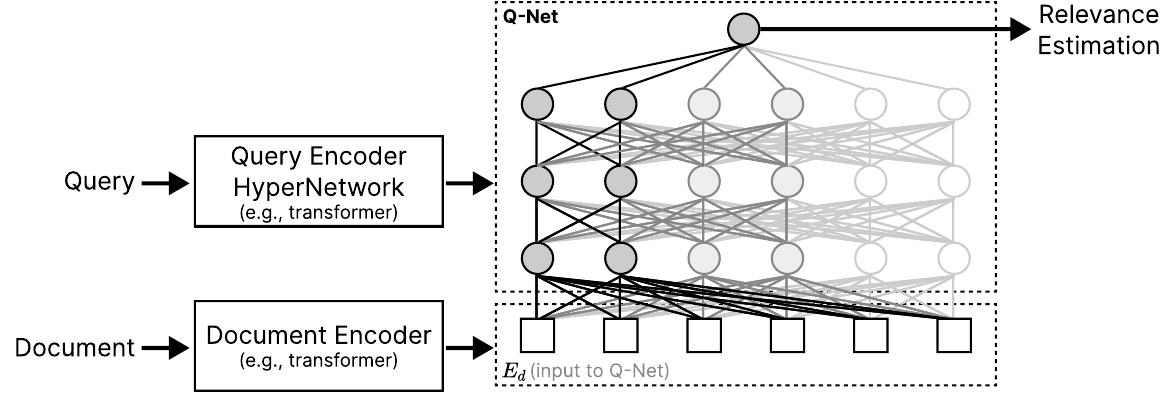}
\caption{Overview of the Matryoshka Hypencoder architecture. Unlike a typical Hypencoder network, the Matryoshka Hypencoder produces Q-Nets that can run at different sizes, which are represented by shades of gray.}
\label{fig:overview}
\end{figure}

In this work, we ask: \textbf{Can we apply the Matryoshka Principle to the Q-Nets produced by a Hypencoder?} To test this, we fine-tune an existing Hypencoder with an adaptation of the Matryoshka loss function~\cite{mrl} to produce Q-Nets of variable sizes (see Figure~\ref{fig:overview}). We find that these smaller Q-Nets exhibit the expected trade-offs between efficiency and effectiveness, enabling a single model to be deployed across multiple operational points. Remarkably, we find that Q-Nets with approximately $7\times$ fewer active parameters provide comparable effectiveness as the full Q-Net in-domain. Keeping half the number of active parameters is required to consistently maintain out-of-domain effectiveness. \sm{We also show that reducing the Q-Net's active parameter count reduces its computational cost.}

In summary, our contributions are as follows. (1) We propose the Matryoshka Hypencoder, a HyperNetwork capable of generating multiple sizes of query-specific relevance networks at once. (2) We show that we can adapt an existing Hypencoder into a Matryoshka Hypencoder through a lightweight fine-tuning process. (3) We show that the resulting model performs well on in-domain and out-of-domain evaluations, allowing for operational cost reductions while maintaining effectiveness. \sm{This work enables more practical deployment settings for the Hypencoder, and the reduced computational cost could allow the retrieval process to run on a CPU instead of GPU (further reducing deployment costs).}

\section{Background and Related Work}

We now cover background on Hypencoders and MRL.

\subsection{Hypencoders}
\label{sec:back:hype}

The Hypencoder is a recent retrieval  proposed by \citet{hypencoder} to addresses the limited expressiveness of linear similarity measurements such as inner products by representing the query as shallow neural network called the Q-Net, resulting in a type of models that balances the efficiency of bi-encoder models and the effectiveness of cross-encoder models

The Hypencoder model consists of a document encoder and a query encoder. The document encoder is a standard BERT-based encoder that produces a single \verb|[CLS]| vector representation $E_d$ for a document, the query encoder consists of a standard BERT-based encoder and an attention-based hyperhead forming a HyperNetwork~\cite{hypernetworks} that is designed and trained to generate the parameters of the Q-Net. In Hypencoders a query $q$ is passed through the query encoder producing embeddings $E_q \in {R}^{n \times h}$, where $n$ is the number of tokens and $h$ is the hidden dimension. The embeddings $E_q$ are passed to the hyperhead generating the weights and biases of the Q-Net. This is implemented using a single-head attention mechanism and a feed-forward layer, producing the parameters of the Q-Net. For each of the $l$ required parameter matrices of the Q-Net, the hyperhead computes  Key and Value matrices and the hidden Representation:
            \begin{equation}
               K_i = E_q W_{K_i} \; \;, \; \; V_i = E_q W_{V_i} \;\; , \;\; H_i = \text{softmax}(\frac{Q_i K_i^T}{\sqrt{h}}) V_i  
            \end{equation}
where $Q_i$ is a set of learnable query embeddings unique to the $i$-th parameter.

The Q-Net is represented as a shallow, multi-layer feed-forward network of a fixed architecture with dynamically generated weights. \citet{hypencoder} uses a 6-layer MLP with LayerNorm and residual connections. For a document embedding $E_d$, the relevance score $s$ is computed by the forward pass of the Q-Net:
\begin{equation} x_0 = E_d  \;\;,\;\;  x_{i+1} = x_i + \text{LayerNorm}(\text{ReLU}(x_i W_i^q + b_i^q))     \end{equation}
where $W_i^q$ and $b_i^q$ are the parameters generated by the hyperhead for query $q$.
         
The results presented in the paper demonstrate that the Hypencoder architecture has outperformed strong bi-encoder baselines on in-domain tasks. More importantly, the performance gap widened substantially on the hard retrieval tasks, validating the strong expressive power of its Q-Nets.

\subsection{Matryoshka Representation Learning (MRL)}
\label{sec:back:mrl}

\looseness -1 While the Hypencoder improves the similarity measurement approach, it introduces a new rigidity, the fixed size of the Q-Net. However, we identify that Matryoshka Representation Learning (MRL)~\cite{mrl} could overcome this rigidity. MRL models produce a single, high-dimensional representation of the input, which can be effectively truncated to a lower-dimensional representation, much like a set of nested Matryoshka dolls (which inspired its name). The approach has also been applied to neural architectures themselves~\cite{DBLP:conf/iclr/LiLL0L25,DBLP:conf/sigir/0032ZKZ25}. These ``2D'' Matryoshka architectures extend the MRL concept along a second dimension: the \textit{depth} of the encoder model. Truncating the model at an earlier layer reduces the number of activated parameters, thereby reducing computational cost.

Their approach trains a single Transformer encoder to produce effective embeddings from not only the final layer but also from various intermediate sub-layers. The total loss is a cumulative sum over both multiple dimensions (width) and multiple layers (depth). While 2DMSE focuses on producing truncatable embedding \textit{vectors} by modifying the encoder's training objective, our work explores a different generalisation of MRL. We focus on creating a truncatable \textit{function} by applying the Matryoshka principle to the parameter space of the Hypencoder's dynamically generated Q-Net.

MRL achieves this by modifying the loss functions. Instead of calculating the loss at the full dimension $d$ the loss is calculated at multiple predefined smaller dimensions $\mathcal{M}$, or granularities (e.g. 128, 256,...etc) by simply truncating the representation vector to the smaller dimensions. The total loss is an aggregation of the losses over all the truncated representations. This forces the model to learn a representation that is well-structured and effective even at smaller dimensionalities. The MRL paper showed that using this training technique led to massive efficiency gains in tasks like classification and retrieval. The standard MRL loss function:

\begin{equation}
    \label{eq:standard_mat_loss}
    \min_{ \{W^{(m)}\}_{m \in \mathcal{M}}, \theta_F } \frac{1}{N} \sum_{i \in [N]} \sum_{m \in \mathcal{M}} c_m \cdot \mathcal{L} \left( W^{(m)} \cdot F(x_i; \theta_F)_{1:m} \, ; \, y_i \right)
\end{equation}

\noindent where $\theta_F$ are the learnable parameters; $W^{(m)}$ is the classifier head for dimension $m$; $\mathcal{M}$ is the set of Matryoshka dimensions; and $c_m$ is the loss weighting coefficient for dimension $m$.

We adapt the principles of MRL from data representations to the relevance function itself. Instead of truncatable embedding vectors, we train a Matryoshka Hypencoder to generate a single set of parameters that can be truncated to form nested Q-Nets of varying widths, enabling flexible efficiency-effectiveness trade-offs.

\section{The Matryoshka Hypencoder (MH)}

We now cover the design of the Matryoshka Hypencoder.

\subsection{Architecture}

This work transformers the original MRL concept from generating a truncatable encoding vector into a truncatable function representation (the parameters of the Q-Net). The objective here is to train the hyperhead to generate a single large set of parameters $\Theta_q^{full}$, for a wide Q-Net which can also be truncated into a subset of these parameters $\Theta_q^{m}$, to form a valid and effective narrower Q-Net for dimension $m \in \mathcal{M}$.

To achieve this, two key modifications are introduced to the baseline Hypencoder architecture. First, architectural changes are implemented to expose and truncate the generated parameters, and the original MarginMSE loss function is modified to become a Matryoshka multi-objective Margin MSE Loss function:\footnote{In a pilot study, we explored several variations combining the Matryoshka from Eq.~\eqref{eq:standard_mat_loss} and standard MarginMSE. This version was the most reliable.}

\begin{equation}
    \label{eq:final_mat_los_compressed}
    \mathcal{L} = \frac{1}{|\mathcal{M}|} \sum_{m \in \mathcal{M}} \left( \left[ s_m(p^+; \Theta_q^{1:m}) - s_m(p^-; \Theta_q^{1:m}) \right] - \Delta s_T(q) \right)^2
\end{equation}
During the training process the hyperhead generates the full set of parameters $\Theta_q^{full}$, for the largest size Q-Net dimension $m_{max}$. The loss function iterates over every Matryoshka dimensions $m \in \mathcal{M}$ truncating the full set of parameters to a smaller parameter set $\Theta_q^{m}$. A temporary Q-Net of a hidden layer width $m$ is assembled using the truncated parameters. Then MarginMSE loss is calculated for this smaller Q-Net. Finally the losses of each dimension $m$ are aggregated according to Eq.~\eqref{eq:final_mat_los_compressed}.

\begin{table}[]
\centering
\caption{Evaluation dataset characteristics.}
\label{tab:dataset}
\begin{tabular}{lcrr}
\toprule
Dataset & In-Domain & \# Queries & \# Docs \\
\midrule
MSMARCO Dev & \cmark & 6,980 & 8.8M \\
DL'19 & \cmark & 43 & 8.8M \\
DL'20 & \cmark & 54 & 8.8M \\
TREC COVID (TC) & \xmark & 50 & 171k \\
FiQA (FQA) & \xmark & 648 & 58k \\
DBPedia (DB) & \xmark & 400 & 4.6M \\
NFCorpus (NF) & \xmark & 323 & 3.6k \\
Webis Touche (WT) & \xmark & 49 & 383k \\
\bottomrule
\end{tabular}
\end{table}

\subsection{Practical Considerations}
\label{sec:finetune}

The training process of the original Hypencoder was very resource-intensive (e.g., 6 days on two A100 GPUs). This presents practical challenges for reproducibility, as well as environmental considerations. Instead of training the Matryoshka Hypencoder in a similar end-to-end fashion, we propose a lightweight training procedure to further fine-tune an existing Hypencoder model into a Matryoshka Hypencoder model. To further improve the training efficiency, we freeze the document and query transformer networks, allowing only the hyperhead of the query encoder to train. \sm{For fair comparisons, we include a version of the Hypencoder further fine-tuned this way, without the Matryoshka modifications.}

\section{Experiments}
\label{sec:exp}

We conduct experiments to answer three research questions about the proposed Matryoshka Hypencoder. We first validate that our efficient fine-tuning procedure described in Section~\ref{sec:finetune} is effective (without applying the Matryoshka Hypencoder architecture):
\begin{enumerate}
\item[\bf RQ1] Can a Hypencoder model, where the document and query encoders are frozen and only the hyperhead is trained, achieve performance comparable to the fully end-to-end trained model?
\end{enumerate}
Then, we run the same fine-tuning procedure under the Matryoshka Hypencoder architecture to answer the following questions:
\begin{enumerate}
\item[\bf RQ2] Does the Matryoshka Principle apply to the space of HyperNetworks under the Matryoshka Hypencoder architecture?
\item[\bf RQ3] Is the Matryoshka Hypencoder model capable of generalising on out-of-domain retrieval tasks?
\item[\bf RQ4] \sm{Are smaller Q-Nets produced by the Matryoshka Hypencoder progressively faster at inference?}
\end{enumerate}

\begin{table}[t]
\centering
\caption{In-domain performance comparison. MH $d$ refers to our Matryoshka Hypencoder at a Q-Net hidden dimension of $d$. Hyp-Fr is our frozen-encoder model. $\dag$ indicates a statistically significant difference ($p < 0.05$) compared to the Original Hypencoder baseline, and * shows significance compared to MH $d=768$.}
\label{tab:in_domain_results}
\resizebox{\columnwidth}{!}{%
\begin{tabular}{l r c c c}
\toprule
\textbf{Model} & \textbf{\# Q-Net} & \textbf{DL'19} & \textbf{DL'20} & \textbf{Dev} \\
& \textbf{Params} & \bf nDCG@10 & \bf nDCG@10 & \bf MRR@10 \\
\midrule
\multicolumn{5}{l}{\textit{Our Matryoshka Hypencoder (MH)}} \\
\hspace{1em}MH $d=768$ & 4,134,912 & .734 & .723 & .380$^\dag$\phantom{*} \\
\hspace{1em}MH $d=512$ & 1,970,176 & .738 & .729 & .381$^\dag$\phantom{*} \\
\hspace{1em}MH $d=256$ & 591,616   & .741 & .728 & .379$^\dag$\phantom{*} \\
\hspace{1em}MH $d=128$ & 197,632   & .734 & .725 & .375$^\dag$* \\
\midrule
\multicolumn{5}{l}{\textit{Baselines}} \\
Hyp-Fr (Ours) & $\approx$4.1M & .728 & .729 & .373$^\dag$\phantom{*} \\
Original Hypencoder~\cite{hypencoder} & $\approx$4.1M & .742 & .730 & .386\phantom{$^\dag$*} \\
BM25 & - & .498 & .479 & .188\phantom{$^\dag$*} \\
\bottomrule
\end{tabular}
}
\end{table}

\subsection{Experimental Setup}
For training all models, we use the pre-processed MS MARCO dataset from Killingback et al.~\cite{hypencoder}, which includes passages scored by a cross-encoder teacher model for knowledge distillation. We set $\mathcal{M}=[128, 256, 512, 768]$\footnote{We also tried training smaller sizes (64 and 32), but the model failed to converge. This observation is in line with prior work that observed challenges when training Matryoshka with a small number of dimensions~\cite{DBLP:conf/sigir/0032ZKZ25}.} for the Matryoshka Hypencoder. Note that the original Hypencoder model used a fixed size of 768.

We evaluate the performance of the model on several standard benchmarks, summarised in Table~\ref{tab:dataset}. For in-domain evaluation, we use the MS MARCO Dev set~\cite{msmarco} and the TREC Deep Learning (DL) 2019 and 2020 passage ranking tracks~\cite{trec_dl_2019, trec_dl_2020}. To assess zero-shot out-of-domain (OOD) generalisation, we use a diverse suite of datasets from the BEIR benchmark~\cite{beir}, including TREC-COVID~\cite{DBLP:journals/sigir/VoorheesABDHLRS20} (biomedical), NFCorpus~\cite{DBLP:conf/ecir/BotevaGSR16} (nutrition), FiQA~\cite{DBLP:conf/www/MaiaHFDMZB18} (financial), DBPedia~\cite{DBLP:conf/sigir/HasibiNXBBKC17} (entity), and Touché~\cite{DBLP:conf/clef/BondarenkoFBGAP20a} (argument retrieval). All evaluation datasets were accessed via the IR Datasets library~\cite{ir_datasets}. This selection of training and evaluation data mirrors the setup in the original Hypencoder paper~\cite{hypencoder}.

We measure the official task metric for each dataset: Mean Reciprocal Rank at rank 10 (MRR@10) for MS MARCO Dev, and 
Normalised Discounted Cumulative Gain at depth 10 (nDCG@10~\cite{DBLP:journals/tois/JarvelinK02}) for the other datasets. We compute statistical significance using a paired t-test and $p<0.05$.

For context, we include the performance of the original Hypencoder model from~\cite{hypencoder} and the performance of PISA's~\cite{DBLP:conf/sigir/MalliaSMS19} implementation of BM25~\cite{DBLP:conf/trec/RobertsonWJHG94}, using its default parameters.

\section{Results and Discussion}

This section answers the research questions posed in Section~\ref{sec:exp}.

\subsection{Frozen-Encoder Training (RQ1)}

Our first research question explores whether a regular Hypencoder can be effectively fine-tuned with frozen transformers from an existing Hypencoder model (as described in Section~\ref{sec:finetune}). Tables~\ref{tab:in_domain_results} and~\ref{tab:ood_results} show the performance of this model (Hyp-Fr), compared with the results from the original Hypencoder. In 7 out of 8 datasets (across both in-domain and out-of-domain evaluations), we observe no statistically significant differences between the results from the original Hypencoder and this variant.
This result demonstrates that the lightweight hyperhead training process is successful.

We answer \textbf{RQ1} in the affirmative: a Hypencoder model can be trained with frozen query and document encoders and achieve performance comparable to that of a fully end-to-end trained model.

\subsection{Architectural Feasibility (RQ2)}

Having established a viable training paradigm, we now evaluate our primary contribution: the Matryoshka Hypencoder.

\begin{table}[t]
\centering
\caption{Zero-shot performance (nDCG@10) on out-of-domain BEIR datasets. TC is TREC-COVID, FQA is FiQA, NF is NFCorpus, DB is DBPedia, and WT is Webis-Touché. $\dag$ indicates a statistically significant difference ($p < 0.05$) compared to the Original Hypencoder baseline, and * shows significance compared to MH $d=768$.}
\label{tab:ood_results}
\resizebox{\columnwidth}{!}{%
\begin{tabular}{l c c c c c}
\toprule
\textbf{Model} & \textbf{TC} & \textbf{FQA} & \textbf{DB} & \textbf{NF} & \textbf{WT} \\
\midrule
\multicolumn{6}{l}{\textit{Our Matryoshka Hypencoder (MH)}} \\
\hspace{1em}MH $d=768$ & .685\phantom{$^\dag$} & .312\phantom{$^\dag$*} & .414 & .324\phantom{$^\dag$} & .272\phantom{*} \\
\hspace{1em}MH $d=512$ & .684\phantom{$^\dag$} & .311\phantom{$^\dag$*} & .414 & .324\phantom{$^\dag$} & .271\phantom{*} \\
\hspace{1em}MH $d=256$ & .678$^\dag$ & .306$^\dag$* & .414 & .324\phantom{$^\dag$} & .266\phantom{*} \\
\hspace{1em}MH $d=128$ & .670$^\dag$ & .304$^\dag$* & .412 & .323\phantom{$^\dag$} & .259* \\
\midrule
\multicolumn{6}{l}{\textit{Baselines}} \\
Hyp-Fr (Ours) & .699\phantom{$^\dag$} & .314\phantom{$^\dag$*} & .419 & .311$^\dag$ & .250\phantom{*} \\
Original Hypencoder~\cite{hypencoder} & .698\phantom{$^\dag$} & .314\phantom{$^\dag$*} & .419 & .324\phantom{$^\dag$} & .258\phantom{*} \\
BM25 & .610\phantom{$^\dag$} & .253\phantom{$^\dag$*} & .303 & .318\phantom{$^\dag$} & .332\phantom{*} \\
\bottomrule
\end{tabular}
}
\end{table}

Table~\ref{tab:in_domain_results} presents the in-domain performance of the Matryoshka Hypencoder model on the TREC DL 2019 and 2020 benchmarks and MSMARCO dev. All of the sizes ($\mathcal{M}=[128,256,512,768]$) exhibit strong retrieval effectiveness, and do not differ from the Original Hypencoder baseline when tested using near-complete relevant assessments (DL'19 and DL'20). However, it performs significantly worse than the Original Hypencoder baseline for the dev set, which contains incomplete relevance assessments. We also see no significant degradations compared to the full-size Matryoshka Hypencoder ($d=768$) for $d=512$ and $d=256$, with only significant degradations for MS MARCO dev at $d=128$.

Meanwhile, we observe substantial reductions in the number of active parameters: $d=512$ needs less than half the number of parameters as the full-sized $d=768$ model, while the reductions for the smaller sizes are more substantial ($7\times$ fewer parameters for $d=256$ and $21\times$ fewer parameters for $d=128$). The non-linear relationship of the reductions is due to the fact that a reduction in the hidden size of the Q-Net's feed-forward network results in a roughly quadratic reduction in the number of parameters for the inner layers, as shown visually in Figure~\ref{fig:overview}. This reduction is important because it allows more documents to be scored in parallel (when using a GPU) and potentially reduces the cost enough to run it on a CPU. \sm{We explore the efficiency of this reduction in RQ4.}

These results answer \textbf{RQ2}: Matryoshka Hypencoder performs well in-domain: at $d=256$ it yields no significant in-domain effectiveness reductions, while using $7\times$ fewer active parameters.

\begin{table}
\centering
\caption{Average document scoring throughput (documents scored per second). The speed up compared to the full-size Q-Net ($d=768$) is given in parentheses.}
\label{tab:query_latency}
\scalebox{0.8}{
\begin{tabular}{lrrrrrr}
\toprule
\bf Q-Net & \bf Dev \phantom{\tiny ($0.0\times$)} & \bf TC \phantom{\tiny ($0.0\times$)} & \bf FQA \phantom{\tiny ($0.0\times$)} & \bf DB \phantom{\tiny ($0.0\times$)} & \bf NF \phantom{\tiny ($0.0\times$)} & \bf WT \phantom{\tiny ($0.0\times$)} \\
\midrule
$d=768$ & 0.9M {\tiny ($1.0\times$)}	& 0.9M {\tiny ($1.0\times$)}	& 0.7M {\tiny ($1.0\times$)}	& 0.9M {\tiny ($1.0\times$)}	& 0.2M {\tiny ($1.0\times$)}	& 0.9M {\tiny ($1.0\times$)} \\
$d=512$ & 1.4M {\tiny ($1.6\times$)}	& 1.3M {\tiny ($1.5\times$)}	& 0.9M {\tiny ($1.3\times$)}	& 1.4M {\tiny ($1.6\times$)}	& 0.2M {\tiny ($1.0\times$)}	& 1.4M {\tiny ($1.6\times$)} \\
$d=256$ & 3.0M {\tiny ($3.4\times$)}	& 2.4M {\tiny ($2.8\times$)}	& 1.4M {\tiny ($2.0\times$)}	& 3.0M {\tiny ($3.4\times$)}	& 0.2M {\tiny ($1.1\times$)}	& 2.7M {\tiny ($3.0\times$)} \\
$d=128$ & 5.6M {\tiny ($6.3\times$)}	& 3.3M {\tiny ($3.9\times$)}	& 1.8M {\tiny ($2.6\times$)}	& 5.6M {\tiny ($6.3\times$)}	& 0.2M {\tiny ($1.1\times$)}	& 4.1M {\tiny ($4.6\times$)} \\
\bottomrule
\end{tabular}
}
\end{table}

\subsection{Out-of-Domain Effectiveness (RQ3)}

We next test how well the Matryoshka Hypencoder generalises to domains and tasks outside of the original training data (web question answering from MS MARCO). Table~\ref{tab:ood_results} presents results on the five BEIR datasets covered by the original Hypencoder paper. The in-domain findings hold across four of the five tasks (TREC-COVID, DBPedia, NFCorpus, and Webis Touche), with no statistically significant degradation compared to $d=768$ for $d=256$ and $d=512$. Meanwhile, significant reductions were observed for FiQA at $d=256$ and $d=128$, suggesting that the lower sizes do lose some generalisation capacity. However, we do note that significant reductions compared to the original Hypencoder model were observed for both TREC-COVID and FiQA at $d=256$ and $d=128$. \sm{At $d=512$, the number of active Q-Net parameters is reduced by half while maintaining comparable effectiveness across all five out-of-domain evaluation sets.}

We answer \textbf{RQ3} in the affirmative: the Matryoshka Hypencoder generalises across domains, however it does lose some generalisation capacity as the size of the Q-Net is reduced.

\subsection{\sm{Efficiency (RQ4)}}
Finally, we investigate the practical utility of our proposed architecture: Are smaller Q-Nets produced by the Matryoshka Hypencoder progressively faster at inference time? To answer this, we measured the average document scoring throughput (number of documents scored per second) of the Q-Net at different sizes on MS MARCO Dev and the five out-of-domain datasets. We only measure the Q-Net scoring time, ignoring the query encoding time, since that is roughly constant across all models and dominated by the transformer encoder part of the network. In all cases, we measure brute-force (exhaustive) retrieval over the collections on a NVIDIA RTX 5000 Ada GPU. We expect that the trends we observe in this setting will translate to throughput measurements when using other hardware (GPUs and CPUs) and when performing approximate search techniques, though we leave this for future work.

The results are presented in Table~\ref{tab:query_latency}. There is a direct and negative correlation between the Q-Net's hidden dimension and the throughput; as its size gets smaller, it becomes faster. This is expected, since a smaller Q-Net requires less computational cost to perform inference. The throughputs are most consistent on the two largest collections: MS MARCO (8.8M documents) and DBPedia (4.6M documents), since the larger samples make average per-document processing time more stable.\footnote{The smaller collections, especially NFCorpus ($3.6k$ documents), had slightly less stable throughput measurements.} They achieve a $6.3\times$ speed up at $d=128$, which is well below the theoretical speed up based on the reduction in parameters ($20.9\times$). This discrepancy is likely due to memory bandwidth limitations. We suspect there may be ways to better optimise scoring, though we leave this for future work. In-domain, similar effectiveness was consistently achieved at $d=256$, which results in a $3.4\times$ speed up. For out-of-domain, $d=512$ consistently yielded similar effectiveness, which gives a speed up of $1.6\times$.

In summary, we answer \textbf{RQ4} affirmatively: smaller Q-Nets are progressively faster to execute. However, we do note that they fall short of the theoretical maximum based on parameter count, so future work could explore further optimising Q-Net inference.

\section{Conclusion}

In this work, we introduced the Matryoshka Hypencoder, a novel architecture that addresses the fixed-complexity limitation of the original Hypencoder. By adapting Matryoshka Representation Learning (MRL) principles to the parameter space of the Q-Net, we successfully trained a single hyperhead to generate nested, variable-width scoring functions. We demonstrated that a lightweight transfer learning approach, fine-tuning a pre-trained model with a multi-objective loss, is a viable and effective training strategy.

Our empirical results show that this architecture achieves an excellent trade-off between efficiency and parameter cost. On in-domain benchmarks, the full-size Matryoshka Q-Net usually performs on par with the original model, while a truncated Q-Net (256 hidden layer size) with $7\times$ fewer parameters achieves statistically comparable performance. Furthermore, we demonstrated that this favourable trade-off is robust, generalising successfully to a diverse suite of out-of-domain, zero-shot retrieval tasks. Crucially, we have shown that the reduction in active parameters translates to a massive gain in practical efficiency. The smaller Q-Nets are progressively faster at inference, with the 128-dimension Q-Net providing up to a $6.3\times$ reduction in query latency on large-scale corpora. This is due to the quadratic reduction in computational cost in the document scoring loop, which is the primary bottleneck in the retrieval process.

We hope that this generalisation of MRL to a function generator inspires more work in the query-specific relevance function space.

\bibliographystyle{ACM-Reference-Format}
\bibliography{references_1}

\end{document}